\pdfoutput=1
\documentclass[prl,twocolumn,showpacs,amsmath,amssymb,floatfix,superscriptaddress]{revtex4-1}

\usepackage{graphicx}%Include figure files
\usepackage{dcolumn}%Align table columns on decimal point
\usepackage{bm}% bold math$V_q$
\usepackage{hyperref}

\begin{document}

\title{Origin of nonlocal resistance in multiterminal graphene on hexagonal-boron-nitride: Fermi surface edge currents rather than Fermi sea topological valley currents}

\author{J. M. Marmolejo-Tejada}
\affiliation{Department of Physics and Astronomy, University of Delaware, Newark, DE 19716, USA}
\affiliation{School of Electrical and Electronics Engineering, Universidad del Valle, Cali, AA 25360, Colombia}
\author{J. H. Garc\'{\i}a}
\affiliation{Catalan Institute of Nanoscience and Nanotechnology (ICN2), CSIC and The Barcelona Institute of Science and Technology,
	Campus UAB, Bellaterra, 08193 Barcelona, Spain}
\author{M. Petrovi\'{c}}
\affiliation{Department of Mathematical Sciences, University of Delaware, Newark,  DE 19716, USA}
\author{P.-H. Chang}
\affiliation{Department of Physics and Astronomy, University of Nebraska Lincoln, Lincoln, Nebraska 68588, USA}
\author{X.-L. Sheng}
\affiliation{Department of Physics and Astronomy, University of Delaware, Newark, DE 19716, USA}
\affiliation{Department of Applied Physics, Beihang University, Beijing 100191, China}
\author{A. Cresti}
\affiliation{Univ. Grenoble Alpes, CNRS, Grenoble INP, IMEP-LaHC, F-38000 Grenoble, France}
\author{P. Plech\'a\v{c}}
\affiliation{Department of Mathematical Sciences, University of Delaware, Newark,  DE 19716, USA}
\author{S. Roche}
\affiliation{Catalan Institute of Nanoscience and Nanotechnology (ICN2), CSIC and The Barcelona Institute of Science and Technology,
	Campus UAB, Bellaterra, 08193 Barcelona, Spain}
\affiliation{ICREA--Institucio Catalana de Recerca i Estudis Avan\c{c}ats, 08010 Barcelona, Spain}
\author{B. K. Nikoli\'{c}}
\email{bnikolic@udel.edu}
\affiliation{Department of Physics and Astronomy, University of Delaware, Newark, DE 19716, USA}

\begin{abstract} The recent observation [R. V. Gorbachev {\em et al.}, Science {\bf 346}, 448 (2014)] of nonlocal resistance $R_\mathrm{NL}$ near the Dirac point (DP) of multiterminal graphene on aligned hexagonal boron nitride (G/hBN)  has been interpreted as the consequence of topological valley Hall  currents carried by the Fermi sea states just beneath the bulk gap $E_g$ induced by the inversion symmetry breaking. However, the valley Hall conductivity $\sigma^v_{xy}$, quantized inside $E_g$, is not directly measurable. Conversely, the Landauer-B\"{u}ttiker formula, as numerically exact approach to observable nonlocal transport quantities, yields $R_\mathrm{NL} \equiv 0$ for the same simplistic Hamiltonian of gapped graphene that generates $\sigma^v_{xy} \neq 0$. We combine {\em ab initio} with quantum transport calculations to demonstrate that G/hBN wires with zigzag edges host dispersive edge states near the DP that are absent in theories based on the simplistic Hamiltonian. Although such edge states exist also in isolated zigzag  graphene wires, aligned hBN is required to modify their energy-momentum dispersion and generate $R_\mathrm{NL} \neq 0$ near the DP persisting in the presence of edge disorder. Concurrently, the edge states resolve  the long-standing puzzle of why the highly insulating state of G/hBN is rarely observed. {\em We conclude that the observed  $R_\mathrm{NL}$ is unrelated to Fermi sea topological valley currents conjectured for gapped Dirac spectra}.
\end{abstract}

%\pacs{72.80.Vp, 73.63.-b, 73.22.Pr, 72.15.Lh, 61.48.Gh}
\maketitle

%Then, for Wannier Hamiltonian, v = 1/hbar * 1.708 eV/k_x = 1.034*10^5 m/s.
% v = 1/hbar * 0.6392 eV/k_x = 3.87*10^4 m/s (lattice constant = 2.504A)
%And plain Graphene, v = 1/hbar * 0.2844 eV/k_x = 1.722*10^4 m/s
%

\begin{figure}
	\includegraphics[scale=0.15,angle=0]{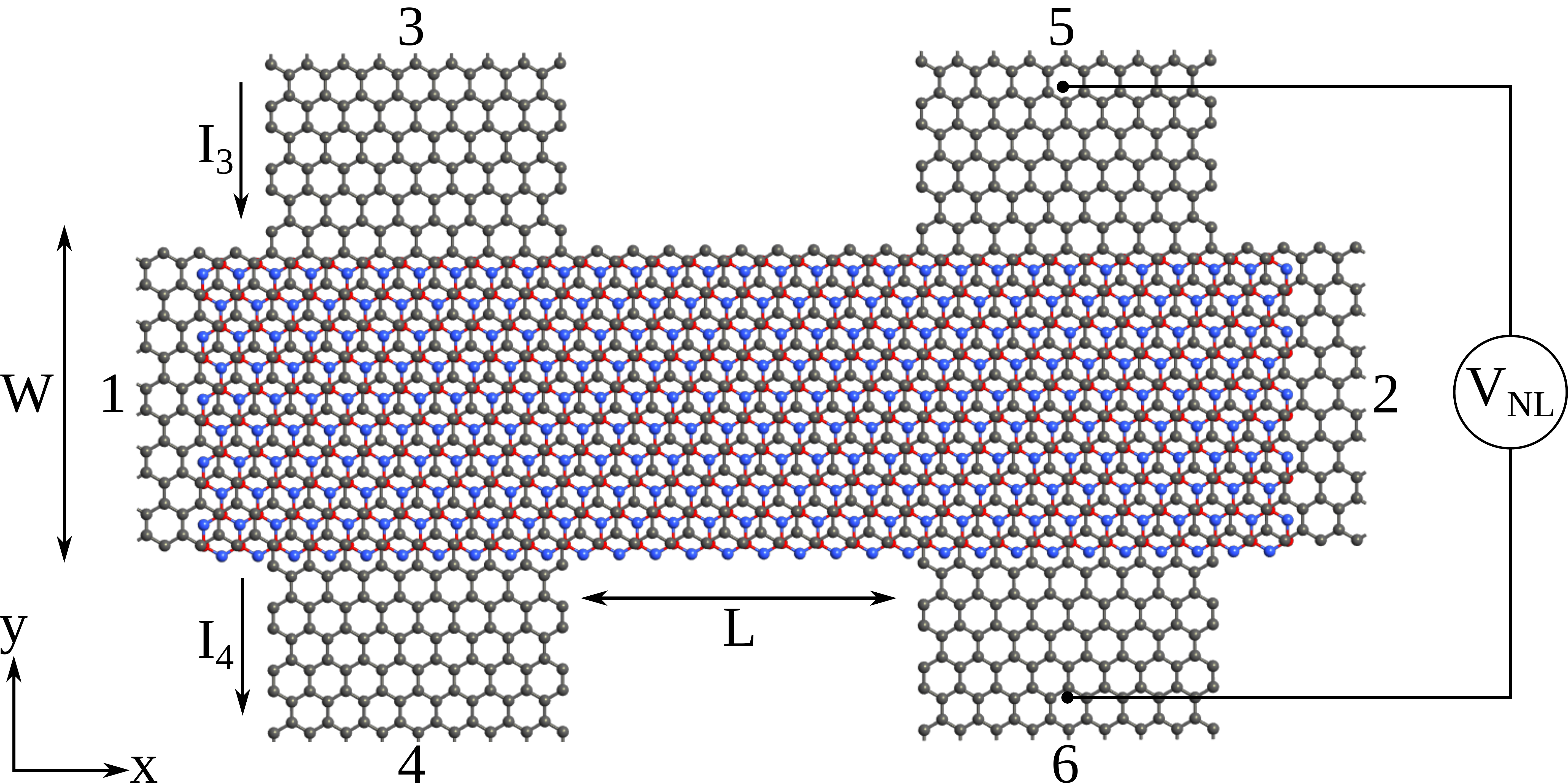}
	\caption{Schematic view of a six-terminal graphene, whose central region is placed onto the hBN substrate, employed in the LB formula calculations  of  nonlocal voltage \mbox{$V_\mathrm{NL}=V_5-V_6$} between leads 5 and 6  and the corresponding nonlocal resistance \mbox{$R_\mathrm{NL}=V_\mathrm{NL}/I_3$} in response to charge current $I_3$. Black, blue and red  circles represent C, N and B atoms, respectively.}
	\label{fig:fig1}
\end{figure}

The recent measurements~\cite{Gorbachev2014} of a sharply peaked nonlocal resistance $R_\mathrm{NL}$ in a narrow energy range near the Dirac point (DP) of multiterminal graphene on hexagonal boron nitride (G/hBN) heterostructures have been interpreted as the manifestation of the valley Hall effect (VHE)~\cite{Xiao2007,Lensky2015,Song2015,Beconcini2016}. In this interpretation, injecting charge current $I_3$ between leads 3 and 4 of the device illustrated in Fig.~\ref{fig:fig1} generates a VH current  in the first crossbar flowing from lead 1 to lead 2, which traverses the channel of length $L$  (\mbox{$\simeq 4$  $\mu$m} in the experiments~\cite{Gorbachev2014}), and it is finally converted into a nonlocal voltage $V_\mathrm{NL}$ between leads 5 and 6  by the inverse VHE in the second crossbar. The corresponding nonlocal resistance \mbox{$R_\mathrm{NL}=V_\mathrm{NL}/I_3$} has been observed previously also near the DP in multiterminal graphene due to an  external magnetic field inducing edge states in the quantum Hall regime or the Zeeman spin Hall effect at higher temperatures~\cite{Abanin2011,Renard2014,Wei2016,Chen2012a,Cresti2016}, as well as due to the spin Hall effect~\cite{Balakrishnan2013,VanTuan2016,Cresti2016}. However, none of these mechanisms is operational in the experiment of Ref.~\cite{Gorbachev2014}. 

Instead, the physics of graphene on hBN with their crystallographic axes {\em aligned} is expected to be governed by the broken spatial inversion symmetry  due to different potentials on two triangular sublattices of carbon atoms induced by the hBN substrate. This opens a gap $E_g$ at the DP of two valleys $K$ and $K'$ in the band structure of an infinite two-dimensional sheet of graphene, where {\em ab initio} calculations have estimated \mbox{$E_g \simeq 58$ meV}~\cite{Giovannetti2007}. In addition, the finite Berry curvature~\cite{Xiao2010} of opposite sign at the two valleys was  predicted to generate valley-dependent transverse conductivities, $\sigma_{xy}^K= e^2/h$ and $\sigma_{xy}^{K'}=- e^2/h$~\cite{Xiao2007,Lensky2015,Song2015}. The VH current is characterized by the transverse VH conductivity, \mbox{$\sigma_{xy}^v=\sigma_{xy}^{K}-\sigma_{xy}^{K'}=2e^2/h$}, and zero transverse charge conductivity, $\sigma_{xy} = \sigma_{xy}^{K}+\sigma_{xy}^{K'} \equiv 0$, within the gap~\cite{Lensky2015,Song2015,Ando2015}. Such charge-neutral valley currents, denoted also as  ``topological''~\cite{Gorbachev2014,Xiao2007,Lensky2015,Song2015} due to the involvement of the Berry curvature hot spots in  reciprocal space, are not conserved but are expected to be long-ranged when the intervalley scattering is weak~\cite{Ando2015}. 

\begin{figure}
	\includegraphics[scale=0.34,angle=0]{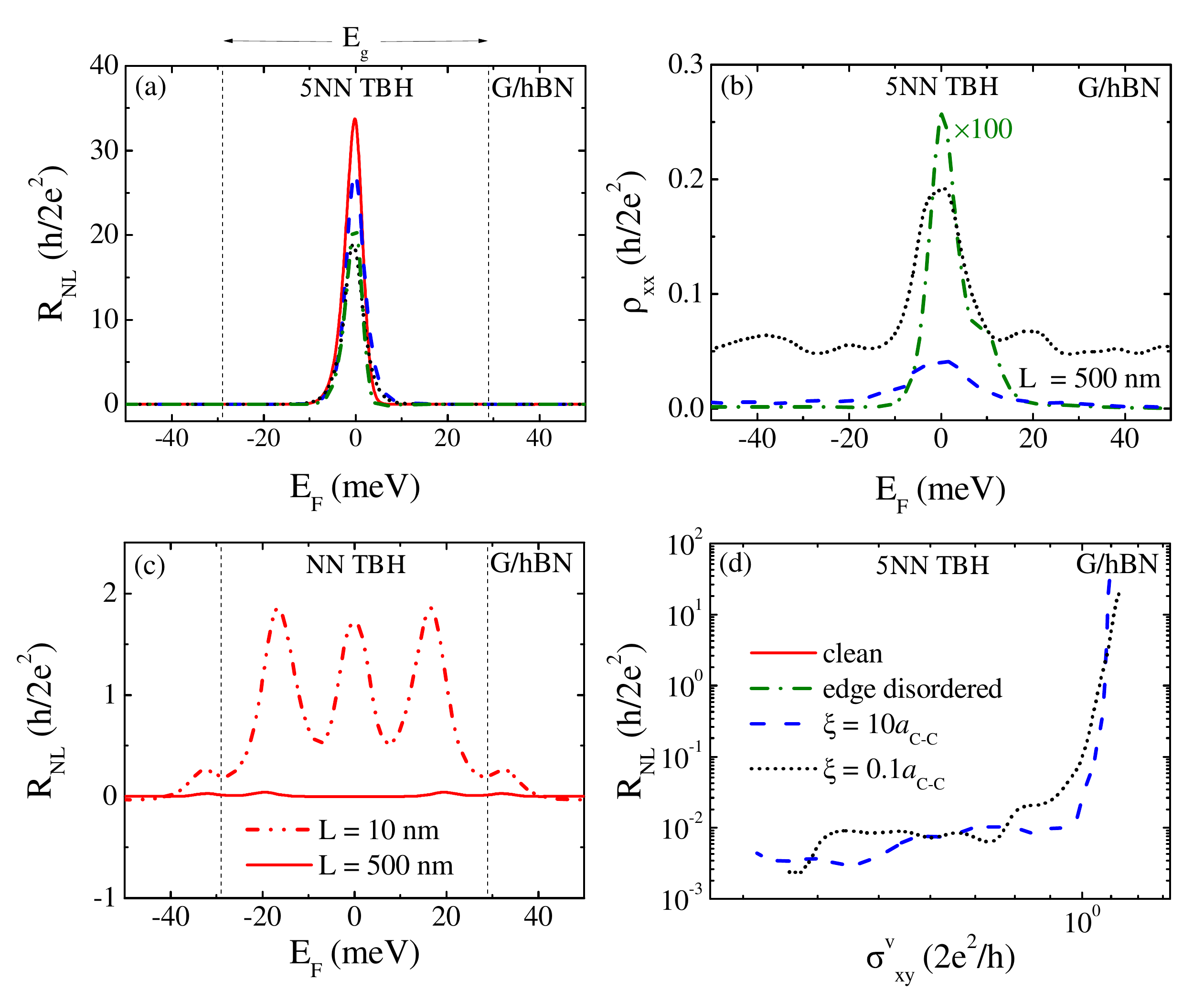}
	\caption{(a),(c),(d) Nonlocal resistance $R_\mathrm{NL}$ and (b) longitudinal resistivity $\rho_{xx}$ at temperature \mbox{$T=20$ K}~\cite{Gorbachev2014} for the device in Fig.~\ref{fig:fig1} whose channel has width \mbox{$W=50$ nm} and length \mbox{$L=500$ nm}, as well as \mbox{$L=10$ nm} in panel (c). We use {\em ab initio} 5NN TBH to describe G/hBN channel with zigzag edges in panels (a),(b),(d); and NN TBH [Eq.~\eqref{eq:tbh}] employed previously~\cite{Gorbachev2014,Xiao2007,Lensky2015,Song2015,Ando2015,Kirczenow2015,Yao2009} in panel (c). The disorder is introduced either by removing carbon atoms along the edges of the channel [Fig.~S5(b) in the SM~\cite{sm}], or by adding impurities into its bulk with 30\% concentration and long-range ($\xi=10 a_\mathrm{C-C}$, where $a_\mathrm{C-C}$ is carbon-carbon bond length, and strength \mbox{$U_p=1.275$ eV} in the model of Sec.~IV in the SM~\cite{sm}) or short-range ($\xi=0.1 a_\mathrm{C-C}$ and effective strength $0.005 \times U_p$) potential. The disorder averaging is performed over 10 samples.}
	\label{fig:fig2}
\end{figure}

However, $\sigma_{xy}^v$ is {\em not directly observable}, and semiclassical transport theories~\cite{Abanin2009} attempting to connect  $\sigma_{xy}^v$ to observable nonlocal resistance~\cite{Gorbachev2014,Song2015,Beconcini2016} 
\begin{equation}\label{eq:rnl}
R_\mathrm{NL} \propto (\sigma_{xy}^v)^2 \rho_{xx}^3 e^{-L/\ell_v},
\end{equation}
are not applicable~\cite{Adam2009} to electronic transport near the DP. In the case of G/hBN, this is further emphasized~\cite{Kirczenow2015} by the presence of the gap forcing electrons to tunnel through the system, which is a phenomenon with no classical analog. The Landauer-B\"{u}ttiker (LB) formalism~\cite{Buttiker1986,Baranger1989,Groth2014}, which offers quantum transport framework to compute $V_\mathrm{NL}$ and $R_\mathrm{NL}$ directly and has been used for decades to model nonlocal transport measurements~\cite{McEuen1990,Roth2009}, yields $R_\mathrm{NL} \equiv 0$ near the DP in Fig.~\ref{fig:fig2}(c) in multiterminal geometries whose channel length is larger than its width ($L/W \simeq 4$ in the experiments of Ref.~\cite{Gorbachev2014}). The numerically exact result in Fig.~\ref{fig:fig2}(c), for a G/hBN system described by the same simplistic Hamiltonian employed~\cite{Gorbachev2014,Xiao2007,Lensky2015,Song2015,Ando2015} to obtain  $\sigma_{xy}^v \neq 0$, is  clearly incompatible with the interpretation of $R_\mathrm{NL} \neq 0$ based on the picture of topological valley currents carried by the Fermi sea states just beneath the gap~\cite{Lensky2015}. Such currents are conjectured to be persistent and circulating in equilibrium~\cite{Lensky2015}, but they become mediative VH currents connecting the two crossbars in Fig.~\ref{fig:fig1} under the application of bias voltage, thereby circumventing the absence of electronic states around the DP while demanding {\em a major overhaul of the LB theory}~\cite{Baranger1989} in which the absence of states within the gap is a fundamental reason for $R_\mathrm{NL} \equiv 0$ in Fig.~\ref{fig:fig2}(c). In geometries with $L<W$, we (as well as Ref.~\cite{Kirczenow2015}) do get $R_\mathrm{NL} \neq 0$ in Fig.~\ref{fig:fig2}(c), but this is trivially explained by transport through evanescent wave-functions able to propagate from first to second crossbar through the gap in such geometries~\cite{VanTuan2016,Kirczenow2015}.

\begin{figure}
	\includegraphics[scale=0.30,angle=0]{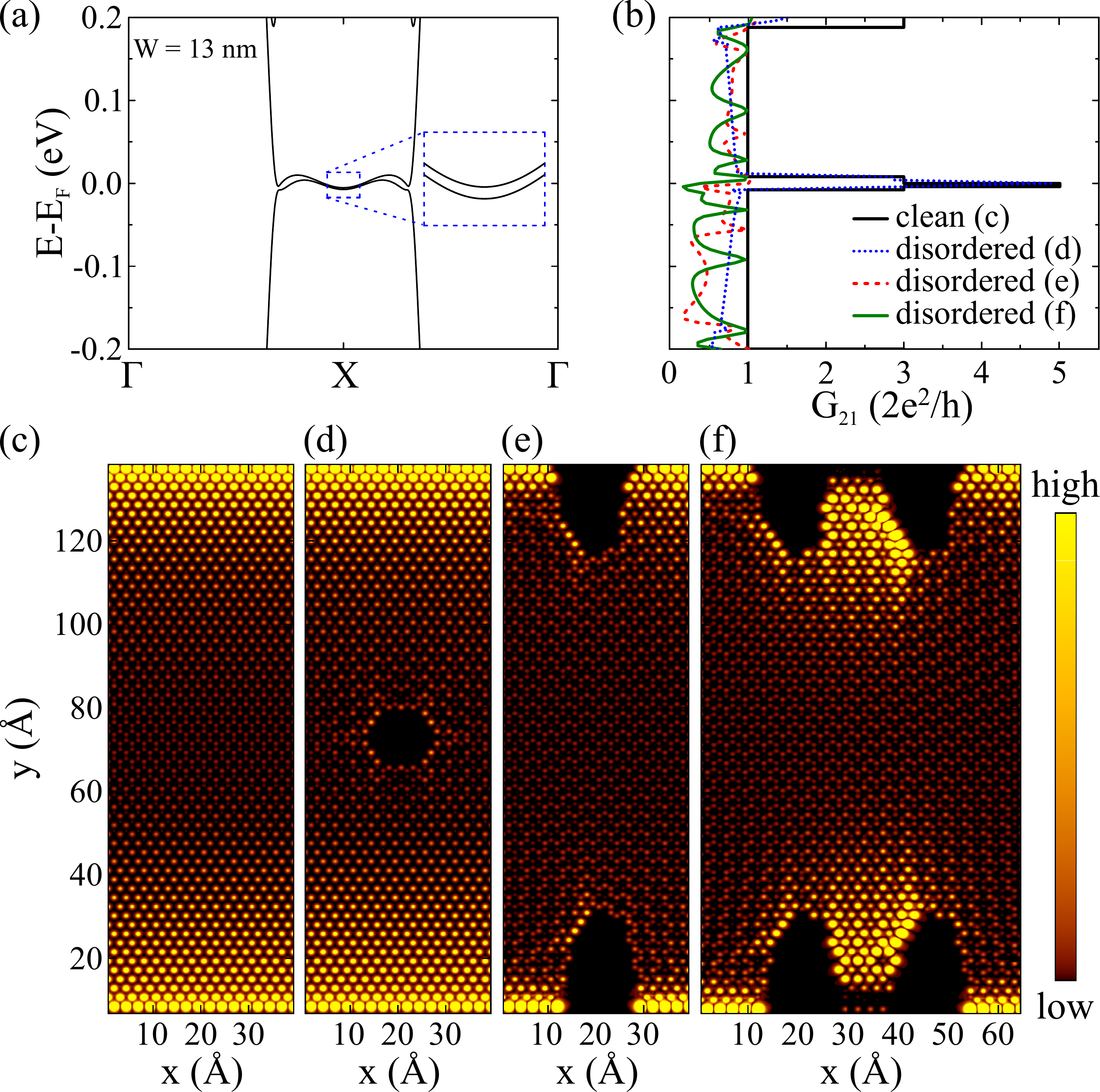}
	\caption{(a) {\em Ab initio} band structure; (b) zero-temperature two-terminal conductance $G_{21}$; and (c)--(f) LDOS at \mbox{$E-E_F=0$} for an infinite G/hBN wire with zigzag edges. The wire is clean in (c), it includes bulk nanopore in (d), or edge disorder in (e) and (f). The wire of width \mbox{$W = 13$ nm} is described by a DFT Hamiltonian, as implemented in the {\tt ATK}~\cite{atk} package, in the basis of double-zeta polarized pseudoatomic orbitals on C, B, N and edge passivating H atoms, and local density approximation is used for the exchange-correlation functional.}
	\label{fig:fig3}
\end{figure}

Another {\em long-standing puzzle} is the metallic-like ordinary longitudinal resistivity,  \mbox{$\rho_{xx} \sim 10$ k$\Omega$}, measured experimentally~\cite{Gorbachev2014} for the G/hBN channel between the two crossbars despite the expected finite gap in its bulk. This suggests the presence of additional conduction pathways, such as edge currents observed very recently~\cite{Zhu2017}, which shunt the highly insulating state at low temperatures. However, previous studies have concluded that edge states are either absent~\cite{Lensky2015} or, when they are present due to edges like zigzag~\cite{Jia2009,Saha2012} or chiral~\cite{Tao2011,Chang2012}, they become gapped near the DP and dispersionless away from it so they do not carry any current~\cite{Yao2009}. The latter conclusion is reproduced in Fig.~S1(f) in the Supplemental Material (SM)~\cite{sm} where we compute band structure of G/hBN wire with zigzag edges using the same simplistic Hamiltonian employed in 
prior studies~\cite{Gorbachev2014,Xiao2007,Lensky2015,Song2015,Ando2015,Kirczenow2015,Yao2009}.

\begin{figure}
	\includegraphics[scale=0.3,angle=0]{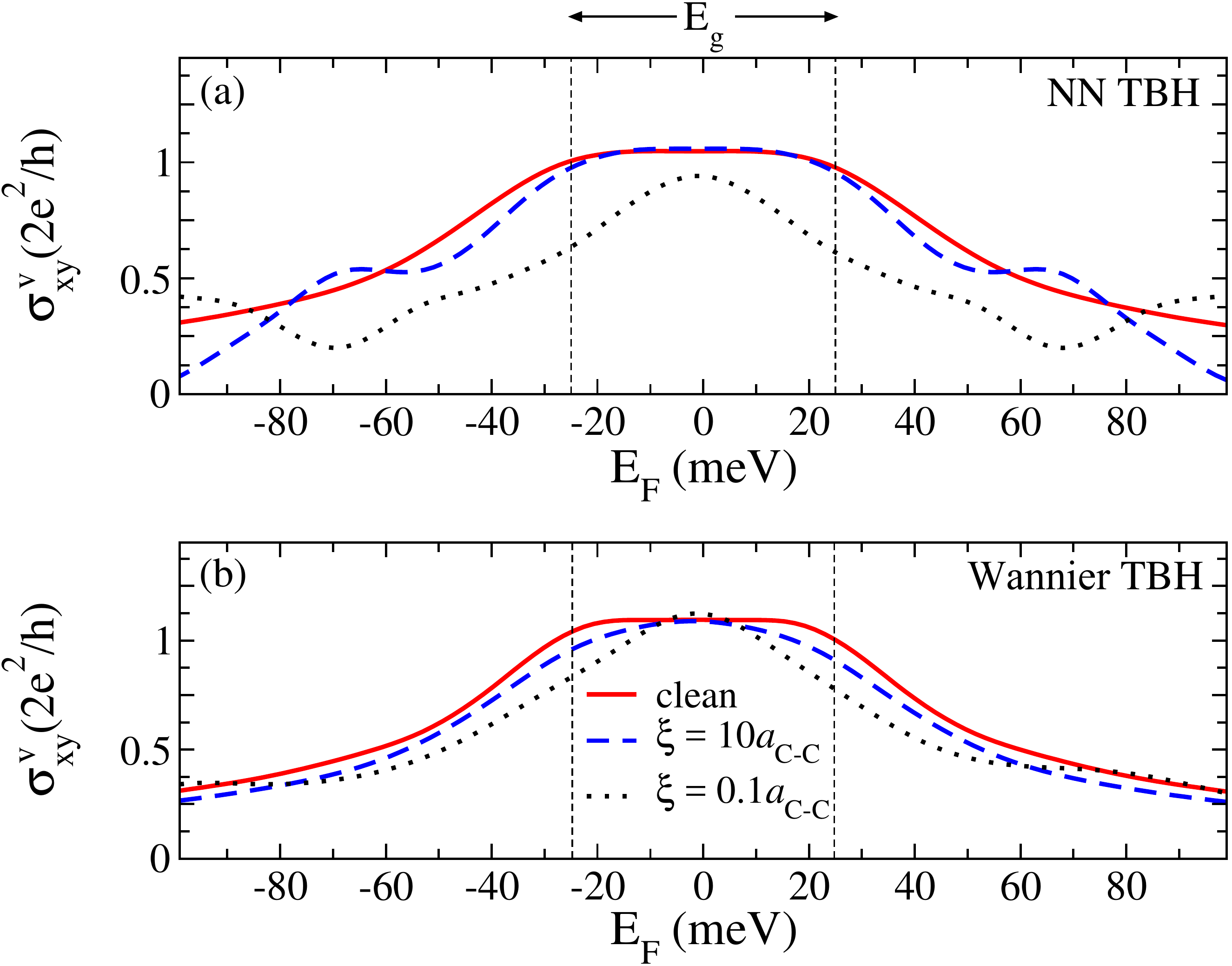}
	\caption{Zero-temperature VH conductivity $\sigma_{xy}^v$ computed by the Kubo formula for a $2048 \times 2048$ G/hBN supercell described by:  (a) NN TBH in Eq.~\eqref{eq:tbh}; or (b) {\em ab initio} Wannier TBH. The properties of long-range and short-range disorder used for dashed and dotted lines, respectively, are the same as in Fig.~\ref{fig:fig2}.}
	\label{fig:fig4}
\end{figure}

In this Letter, we resolve {\em both puzzles}---\mbox{$R_\mathrm{NL} \equiv 0$} [Fig.~\ref{fig:fig2}(c)] obtained for multiterminal G/hBN whose bulk counterpart exhibits $\sigma^v_{xy} \neq 0$ [Fig.~\ref{fig:fig4}(a)]; and metallic-like $\rho_{xx}$ despite presumed gap opening around the DP---by performing density functional theory (DFT) calculation combined with quantum transport simulations, based on both the multiterminal LB formula (Figs.~\ref{fig:fig2} and ~\ref{fig:fig3}) and the Kubo formula (Fig.~\ref{fig:fig4}). We demonstrate that these puzzles are an artifact of the simplistic Hamiltonian~\cite{Gorbachev2014,Xiao2007,Lensky2015,Song2015,Ando2015,Kirczenow2015,Yao2009} which inadequately describes G/hBN  wires. For example, in contrast to previously obtained~\cite{Yao2009} gapped band structure for all types of G/hBN wires, the {\it ab initio} band structure [Fig.~\ref{fig:fig3}(a)]  of G/hBN wires with zigzag edges has no gap. The {\it ab initio} Hamiltonian combined with the LB formula yields $R_\mathrm{NL}$ [Fig.~\ref{fig:fig2}(a)] and $\rho_{xx}$ [Fig.~\ref{fig:fig2}(b)] whose features closely match those measured in Ref.~\cite{Gorbachev2014}: ({\em i}) $\rho_{xx}$ peak is wider than $R_\mathrm{NL}$ peak whose height decays exponentially with increasing $L$ (Fig.~S6(b) in the SM~\cite{sm});  ({\em ii}) aligned hBN is {\em necessary} to obtain $R_\mathrm{NL} \neq 0$ in Fig.~\ref{fig:fig2}(a)---isolated zigzag graphene wire also has edge states (Fig.~S3(a)--(f) in the SM~\cite{sm}), but its band structure leads to $R_\mathrm{NL} \equiv 0$ (Fig.~S4 in the SM~\cite{sm}). 

Since the resolution of the two puzzles relies on an accurate  Hamiltonian for G/hBN wires, as well as its combination with a proper quantum  theory of observable transport quantities, we first summarize inconsistencies arising in previous theoretical analyses.  The seminal arguments~\cite{Xiao2007} for the VHE  are based on semiclassical transport theory describing the motion of a narrow wave-packet constructed by superposing~\cite{Xiao2007,Puetter2012} eigenstates of gapped Dirac Hamiltonian $\hat{H}_\mathrm{D}$ with dispersion $\epsilon_\mathbf{k}$. This assumes that each valley of  G/hBN can be described by  \mbox{$\hat{H}_\mathrm{D} = \hbar v_F (\hat{\bm \sigma} \cdot \hat{\mathbf{k}}) + \Delta \hat{\sigma}_z$}, where $v_F$ is the Fermi velocity, $\hat{\bm \sigma}=(\hat{\sigma}_x,\hat{\sigma}_y)$ is the vector of the Pauli matrices corresponding to the sublattice degree of freedom and $\hbar \hat{\mathbf{k}}$ is the momentum operator. The wave-packet velocity~\cite{Xiao2010},  $\mathbf{v}_\mathbf{k} = \frac{1}{\hbar} \partial \epsilon_\mathbf{k}/\partial \mathbf{k} + d\mathbf{k}/dt \times {\bm \Omega}_\mathbf{k}$,  acquires an anomalous term due to the Berry curvature ${\bf \Omega}_\mathbf{k}$ hot spot near the apex of the valley described by $\hat{H}_\mathrm{D}$. Since ${\bm \Omega}_\mathbf{k}$ in valley $K$ has opposite direction to that in valley $K'$, electrons belonging to two valleys will be separated~\cite{Puetter2012} in the opposite transverse directions in the presence of an applied electric field $\mathbf{E}$ which is required to accelerate electrons according to $\hbar d\mathbf{k}/dt = e \mathbf{E}$. This gives rise~\cite{Xiao2007} to \mbox{$\sigma_{xy}^{K,K'}=\frac{e^2}{\pi h} \int\!\! d^2k \,  f(\varepsilon_\mathbf{k}) \Omega_\mathbf{k}$}, where $f(\varepsilon_\mathbf{k})$ is the Fermi function forcing the integration over the whole Fermi sea, i.e., from the bottom of the band to the Fermi level $E_F$.

However, it has already been pointed out in Ref.~\cite{Kirczenow2015} that nonzero  $\mathbf{E}$ {\em cannot} appear in the linear-response limit of the multiterminal LB formula~\cite{Buttiker1986,Baranger1989,Groth2014}. The experiments measuring $R_\mathrm{NL}$ are carefully kept~\cite{Gorbachev2014} in the linear-response regime in order to avoid heating of the device and the ensuing thermoelectric effects that can add large spurious contributions to $R_\mathrm{NL}$~\cite{Renard2014}. The multiterminal LB formula, \mbox{$I_p = (2e^2/h) \sum_q G_{pq} (V_p - V_q)$}, relates the total charge current $I_p$ in lead $p$ to voltages $V_q$ in all other leads {\em via} the conductance coefficients \mbox{$G_{pq} = \int\!\! dE\, (-\partial f/\partial E) T_{pq}(E)$}, where the derivative of the Fermi function confines the integration to a shell of states of width $\sim k_BT$ around $E_F$~\cite{Baranger1989}. The transmission functions $T_{pq}(E)$ do not include any effect of $\mathbf{E}$~\cite{Buttiker1986,Baranger1989,Groth2014}. We use the multiterminal LB formula implemented in {\tt KWANT} package~\cite{Groth2014,kwant} to compute $V_\mathrm{NL}$ and  $R_\mathrm{NL}$ in response to an injected current $I_3=-I_4$~\cite{VanTuan2016} while keeping $I_p \equiv 0$ in the other four leads. The same procedure allows us to compute $\rho_{xx}=R_{4\mathrm{T}}W/L$ from the four-terminal resistance $R_{4\mathrm{T}}=(V_3-V_4)/I_1$ obtained by injecting current $I_1=-I_2$ into the device in Fig.~\ref{fig:fig1} and by imposing voltage probe condition, $I_p \equiv 0$, in leads $p=3$--$6$. 

The semiclassical arguments for the origin of $\sigma_{xy}^v$ can be replaced by quantum transport theory based on the Kubo formula~\cite{Ando2015,Bastin1971}, which requires to first  obtain~\cite{Ferreira2015,Garcia2015} the velocity operator $\hat{\mathbf{v}}$. The physical consequences of the equation of motion for $\hat{\mathbf{v}} = v_F \hat{\bm \sigma}$ defined by $\hat{H}_\mathrm{D}$~\cite{Ando2015}, $d\mathbf{v}/dt = \frac{1}{i \hbar} [\hat{\mathbf{v}}, \hat{H}_\mathrm{D}] = 2v_F^2(\hat{\mathbf{k}} \times \hat{\bm \sigma}) - \frac{E_g}{\hbar} \hat{\mathbf{v}} \times \mathbf{e}_z$, are extracted by finding its expectation value  in a suitably prepared wave-packet~\cite{Puetter2012,Nikolic2005c} injected with initial velocity into G/hBN where it can propagate even in the absence of any external electric field~\cite{Nikolic2005c}. The first term of $d\mathbf{v}/dt$ causes {\em Zitterbewegung} motion of the center of wave-packet, while the second one acts on it like a Lorentz force due to an effective magnetic field in the direction of the unit vector $\mathbf{e}_z$ perpendicular to the graphene plane. For electrons from the $K'$ valley, $\hat{v}_y = - v_F \hat{\sigma}_y$, leading to opposite direction of such force.

The gapped Dirac Hamiltonian $\hat{H}_\mathrm{D}$ is a long wavelength limit of a more general tight-binding Hamiltonian (TBH) which accounts for both valleys and can, therefore, be used  to capture intervalley scattering effects. The TBH, which is preferred for numerical calculations using the LB~\cite{Kirczenow2015} or the Kubo formula (as well as for simulations of wave-packet dynamics~\cite{Puetter2012}), is defined on a honeycomb lattice hosting a single $p_z$ orbital per site
\begin{equation}\label{eq:tbh}
\hat{H}_\mathrm{TB} = \sum_i \varepsilon_i \hat{c}^\dagger_i \hat{c}_i - t_1 \sum_{\langle ij \rangle} \hat{c}^\dagger_i \hat{c}_j.
\end{equation}
Here $\hat{c}^\dagger_i$  ($\hat{c}_i$) creates (annihilates) an electron at site $i$; hopping \mbox{$t_1=2.7$ eV} is nonzero between  nearest-neighbor (NN) carbon atoms; and  the on-site energy $\varepsilon_i=\pm \Delta$, responsible for the gap $E_g=2\Delta$, is positive on one atom of the graphene unit cell and negative on the other one to take into account the staggered potential induced by the hBN substrate (while neglecting any reorganization of chemical bonding or change in the atomic order of graphene and hBN layers). We use \mbox{$\Delta= 29$ meV}~\cite{Giovannetti2007} to compute $R_\mathrm{NL}$ in Fig.~\ref{fig:fig2}(c) for G/hBN  channel with zigzag edges, as well as to compute its band structure (Fig.~S1(f) in the SM~\cite{sm}) exhibiting gapped flat bands~\cite{Yao2009}.   

Figure~\ref{fig:fig4}(a) shows $\sigma_{xy}^v$, computed using the Kubo formula~\cite{Bastin1971,Garcia2015}  combined with a valley-projection scheme~\cite{Cresti2016,sm,Settnes2017}, for a rhomboid supercell of G/hBN with periodic boundary conditions  described by $\hat{H}_\mathrm{TB}$ in Eq.~\eqref{eq:tbh}. The supercell is either clean or it contains long- or short-range disorder (Sec. IV in the SM~\cite{sm}) as additional terms in the on-site energy $\varepsilon_i$. In the clean limit, we confirm~\cite{Ando2015} that  $\sigma_{xy}^v=2e^{2}/h$ is quantized inside the gap [Fig.~\ref{fig:fig4}(a)], as well as that the Fermi sea states just beneath the gap provide the main contribution to it~\cite{Lensky2015}. For long-range disorder that does not mix valleys, $\sigma_{xy}^v$  remains close to the clean limit  within a smaller energy range than $E_g$ [Fig.~\ref{fig:fig4}(a)] due to disorder-induced broadening of the states~\cite{Ando2015}. High concentration of valley mixing short-range disorder slightly reduces $\sigma_{xy}^v$ [Fig.~\ref{fig:fig4}(a)].  

In order to replace $\hat{H}_\mathrm{TB}$ with a more accurate Hamiltonian, we proceed by computing the {\em ab initio} band structure of G/hBN wires using local-orbital pseudopotential DFT, as implemented in {\tt ATK}~\cite{atk} and {\tt OpenMX}~\cite{openmx,Ozaki2003} packages, whose Kohn-Sham Hamiltonian  can be also easily combined with the LB formula~\cite{Brandbyge2002}.  We assume stacking where one C atom is over a B atom and the other C atom in the unit cell  is centered above the hBN hexagon, as energetically most stable configuration found in DFT calculations~\cite{Giovannetti2007}. The DFT Hamiltonian for a wire---composed of  C, B and N atoms, as well as H atoms passivating dangling bonds along the zigzag edges---produces the gapless band structure in Fig.~\ref{fig:fig3}(a) and the corresponding  zero-temperature two-terminal conductance $G_{21}=5 \times 2e^2/h$ [Fig.~\ref{fig:fig3}(b)] near the DP (at $E-E_F=0$). This value is insensitive to the bulk nanopore [Fig.~\ref{fig:fig3}(d)], signifying edge transport, but it is reduced to $G_{21} \lesssim 2e^2/h$ in the presence of edge vacancies [Fig.~\ref{fig:fig3}(e)] or quantum interferences generated by edge vacancies in series [Fig.~\ref{fig:fig3}(f)] due to lack of  topological protection against backscattering~\cite{Yao2009,Li2010}. The valley-polarized states~\cite{Rycerz2006} above and below the DP are bulk states since $G_{21}$ is reduced at those energies by the presence of the bulk nanopore [Fig.~\ref{fig:fig3}(d)]. The edge state are visualized by plotting the local density of states (LDOS) in Figs.~\ref{fig:fig3}(c)--(f), which is peaked near the edges but it remains nonzero in the bulk~\cite{Chang2012}. This can be  contrasted with topologically protected edge states in quantum (ordinary~\cite{Cresti2016}, spin~\cite{Chang2014} and anomalous~\cite{Sheng2017}) Hall insulators where LDOS in the bulk vanishes. The corresponding local current distributions (Fig.~S5 in the SM~\cite{sm}) shows that edge currents can survive edge disorder breaking G/hBN wire into short zigzag-edge segments, as suggested also by experiments~\cite{Zhu2017}.

Since the usage of the full DFT Hamiltonian is prohibitively expensive for large number of atoms ($\sim 10^6$ in our LB or Kubo formula calculations), we derive simpler {\em ab initio} TBHs~\cite{Fang2015}. A widely-used approach for this purpose is to transform the DFT Hamiltonian to a basis of maximally localized Wannier functions in a selected energy window around $E_F$~\cite{Marzari2012}. We combine Wannier TBH (truncated~\cite{Fang2015} to third NN hoppings without loss of accuracy in the energy window considered in Fig.~\ref{fig:fig4}) with the  Kubo formula in Fig.~\ref{fig:fig4}(b) where we find quantized $\sigma_{xy}^v$ in the gap, as well as its surprising resilience to short-range disorder that was not exhibited in Fig.~\ref{fig:fig4}(a) for the simplistic TBH in Eq.~\eqref{eq:tbh}. However, the Wannier TBH applied to G/hBN wires generates much larger group velocity $\partial \varepsilon_{k_x}/\hbar \partial k_x$ of edge state bands near the DP (Fig.~S1(d) in the SM~\cite{sm}) than in Fig.~\ref{fig:fig3}(a) due to lack of information about atoms (like H) passivating bonds of edge carbon atoms. Nevertheless, such  Wannier TBH could be useful for fully encapsulated G/hBN wires where edges are not exposed to the environment~\cite{Gorbachev2014,Zhu2017}. 

Therefore, we derive an alternative {\em ab initio} TBH that precisely fits (Fig.~S1(c) in the SM~\cite{sm}) the bands around the DP in Fig.~\ref{fig:fig3}(a). This requires up to the fifth NN hoppings and on-site energies in Eq.~\eqref{eq:tbh}, with adjusted values along the edges (Fig.~S2 in the SM~\cite{sm}). The {\em ab initio} 5NN TBH combined with the multiterminal LB formula leads to  $R_\mathrm{NL} \neq 0$  exhibiting sharp peak [Fig.~\ref{fig:fig2}(a)] near the DP in the case of ballistic G/hBN channel. The introduction of disorder---edge (Fig.~S5(b) in the SM~\cite{sm}) or bulk 
of different range (in the model of Sec.~IV in the SM~\cite{sm})---reduces the height of $R_\mathrm{NL}$ peak [Fig.~\ref{fig:fig2}(a)], while concurrently generating peak of $\rho_{xx}$ [Fig.~\ref{fig:fig2}(b)]. Although metallic-like $\rho_{xx}$ could arise due to trivial reasons, such as charge inhomogeneity induced by chemical or electrostatic doping, recent  imaging~\cite{Zhu2017} of proximity-induced supercurrents flowing near the edges of G/hBN wires is in accord with LDOS [Fig.~\ref{fig:fig3}(c)--(f)] or local current distributions (Fig.~S5 in the SM~\cite{sm}). 

In conclusion, $R_\mathrm{NL}$ peak exists in the clean limit [where $\rho_{xx} \rightarrow 0$] and is reduced by the disorder, whereas $\sigma_{xy}^v$ is insensitive to disorder at the DP [Fig.~\ref{fig:fig4}(b)]. Furthermore, plotting $R_\mathrm{NL}(E_F)$ vs. $\sigma_{xy}^v(E_F)$ in Fig.~\ref{fig:fig2}(d), computed for a range of $E_F \ge 0$ independently while using the same long-range or short-range disorder in both the LB and Kubo formula calculations, reveals that  these two quantities are not related  in a way conjectured by Eq.~\eqref{eq:rnl}. This strengthens our principal conclusion---the observed  $R_\mathrm{NL}$ is driven by particular gapless band structure and its edge eigenstates carrying current near the DP, rather than by the Fermi sea dissipationless topological valley currents conjectured for the gapped Dirac spectra.

\begin{acknowledgments}
We thank I. V. Borzenets and K. Komatsu for insightful discussions. J. M. M.-T., X.-L. S. and B. K. N. were supported by NSF Grant No. CHE 1566074. J. M. M.-T. also acknowledges support from COLCIENCIAS of Colombia. M.~P. and P.~P. were supported by ARO MURI Award No. W911NF-14-0247. J. H. G. and S. R. were supported by: the European Union’s Horizon 2020 research and innovation programme (Grant No. 696656); the Spanish Ministry of Economy and Competitiveness and the European Regional Development Fund [Project No. FIS2015-67767-P (MINECO/FEDER)]; and by the Secretaria de Universidades e Investigaci\'{o}n del Departamento de Econom\'{i}a y Conocimiento de la Generalidad de Catalu\~{n}a, and the Severo Ochoa Program (MINECO, Grant No. SEV-2013-0295).  The supercomputing time was provided by XSEDE, supported by NSF Grant No. ACI-1053575, and by the Barcelona Supercomputing Center (Mare Nostrum), under PRACE Project No. 2015133194.
\end{acknowledgments}

% as well as their resilience to disorder, enhanced by hBN substrate, despite zero topological invariants (like Chern number or $\mathbb{Z}_2$) in the bulk band structure;
%********************references************************************************************************

%BibTeX
%Windows:
%\bibliographystyle{C:/BIBTEX/prsty}
%\bibliography{C:/BIBTEX/qttg}

%Linux:
%\bibliographystyle{apsrev}
%\bibliography{$HOME/TEX/BIBTEX/qttg}

\end{document}